# Protocols for Kak's Cubic Cipher and Diffie-Hellman Based Asymmetric Oblivious Key Exchange

Abhishek Parakh

**Abstract**

*This paper presents protocol for Kak's cubic transformation and proposes a modification to Diffie-Hellman key exchange protocol in order to achieve asymmetric oblivious exchange of keys. The cubic transformation may be used to develop a cryptosystem that is more efficient than Rabin's cryptosystem. The oblivious key exchange has applications in mutual exchange of secrets and $m$-out-of-$n$ oblivious transfer.*

## 1. Introduction

It was shown in [1] that the cubic transformation could be used for public-key applications and digital signatures, in which message $m$ is encrypted to cipher $c$ using $c = m^3 \bmod n$, $n = pq$ and $p$ and $q$ are primes, and $\phi(n)$ Euler's totient function is not relatively prime to 3. Earlier, in his study of the square transformation, Rabin [2] had discounted the cubic transformations, having implicitly taken $n$ to be divisible by 9. Kak's cubic transformation deals with the case that Rabin overlooked, i.e. a cubic transformation modulo $n = pq$, where $\phi(n) = (p-1)(q-1)$ is divisible by 3 but not by 9.

This paper presents a protocol for the implementation of the cubic transformation and also presents a method for its use in oblivious transfer. It is organized as follows: Section 2 summarizes the mathematics of the cubic transformation, and then we present a protocol for it. Section 3 deals with the asymmetric nature of oblivious transfer arising from the cubic transformation. We then discuss a modification to Diffie-Hellman key exchange protocol in order to achieve an oblivious transfer of keys between two parties.

## 2. Cubic transformation modulo prime $p$

Consider $c = m^3 \bmod p$, where $p = 3k+1$ is a prime and $p = 3 \bmod 4$. Since $p-1$ and 3 have at least one common divisor (namely 3), the transformation is not one-to-one. Thus three different values of $m$ will map to the same $c$. If $\phi(p)$ is not divisible by 9, we can obtain one of the three values of $c^{\frac{1}{3}}$ by the following operations:

$$c^{\frac{1}{3}} = \begin{cases} c^{\frac{p+2}{9}}, & \text{if the sum of digits of } \phi(p) \text{ is divisible by 6 and} \\ c^{\frac{2p+1}{9}}, & \text{if the sum of digits of } \phi(p) \text{ is divisible by 3 and not by 6.} \end{cases}$$

An inverse by exponentiation will not exist if $\phi(p)$ is divisible by 9. The cube roots of 1 are 1, $\alpha$ and $\alpha^2$. The value of $\alpha$ can be calculated as follows:

$$\alpha = \frac{-1 + (p-3)^{\frac{p+1}{4}}}{-1 - (p-3)^{\frac{p+1}{4}}}$$

The value of $\alpha$ can be used to find all the cubic roots of a number, once one of them has been obtained.

## 2.2 Cubic transformation modulo a composite $n$

Consider the cubic transformation modulo $n = pq$ and $p$ and $q$ are primes and $\phi(n) = (p-1)(q-1)$ is divisible by 3 but not 9. Then the inverse is calculated as:

$$c^{1/3} = \begin{cases} c^{\frac{\phi(n)+3}{9}}, & \text{if the sum of digits of } \phi(n) \text{ is divisible by 6 and} \\ c^{\frac{2\phi(n)+3}{9}}, & \text{if the sum of digits is divisible by 3 and not by 6.} \end{cases}$$

Before we describe the protocol, the user may take note that the following information is public:
1. $n$, however factors of $n$ (namely $p$ and $q$) are private.
2. $\alpha$ is public.

## 2.3 The Protocol

We present the protocol for the cubic transformation with the help of example given in [1]. Consider a transformation modulo a prime number viz. $c = m^3 \bmod 31$. Computing $\alpha$ we get 1, 5 and 25. Suppose the transmitter (Bob) and receiver (Alice) agree upon $\alpha = 5$. This information can be published on websites and made public. Let Bob choose message $m = 7$. He computes $m, m\alpha, m\alpha^2$ which turn out to be 4, 20 and 7 respectively. Now Bob rearranges them in order 4, 7, 20 and assigns then ranks 1, 2 and 3.

We define the ranks in binary format as follow:
Rank 1 = 0
Rank 2 = 01
Rank 3 = 11

Bob transmits the message $m = 7$ applying cubic transformation and appending the encrypted message with rank bits. This is done for each message, one at a time.

Alice reads the incoming bits starting from the least significant bit (LSB). The rank information is to be decoded as follows:
1. If Alice encounters 0, she stops reading further. Now, Alice knows the message corresponds to rank 1.
2. If she encounters 1 in the first LSB, she reads the second LSB. If the second LSB is 0 the rank is 2 and if the second LSB is 1, the rank is 3.

In our example Bob sends $c = 7^3 \bmod 31 = 2$, appended with the rank information 2. Thus the complete message in binary format is:

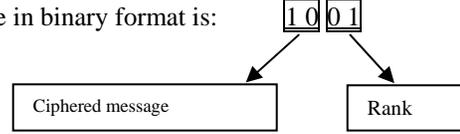

We now consider the operations in binary form. Upon receiving 1001, Alice starts reading from the first LSB. She finds a 1, so she proceeds to read the second LSB which is a 0. Now, Alice knows the rank is 2. She then separates the message from the rank bits as Ciphered message =10 and Rank = 01.

Now converting back to decimal form, Alice deciphers the message 2 (equivalent to 10 in binary form) as $2^7 \bmod 31 = 4$. The other cube roots will be 4 x 5 = 20 mod 31 and 20 x 5 = 7 mod 31. The rank information which Alice already has helps her select $m = 7$. The same protocol can be applied to the case of cubic transformation modulo a composite $n = pq$.

Now we consider the application of the above protocol to the public key cryptography system. Alice generates a private key $KR_A$ and a public key $KU_A$. Bob wishes to send a message $m$ to Alice. The sequence of steps is depicted in figure 1.

In Figure 1, $r$ is the rank information and || means appending the rank information to the ciphered message $m$. $D$ and $E$ are the decryption and encryption methods respectively. The $s$ function separates the ciphered message from the rank information and decodes the rank information. The rank information $r$ can be sent back to Bob as an acknowledgment.

Thus the sequences of steps to be followed are:
1. Alice publishes its public key $KU_A$, $n$ and $\alpha$.
2. Bob encrypts the message using Alice's public key and Alice's encryption method.
3. Bob then appends the rank information to the ciphered message and transmits to Alice.
4. Alice decodes the rank information and separates if from the ciphered message.
5. Alice deciphers the message with the help of the rank information and sends back the rank to Bob as an acknowledgement.

If needed Bob can verify this acknowledgement to make sure that Alice had not made an error.

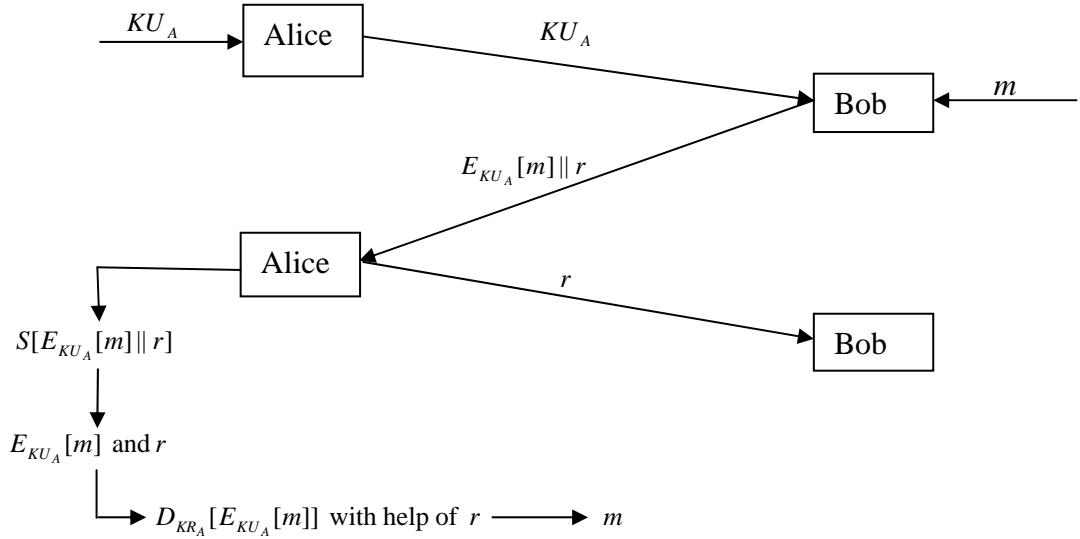

**Figure 1. Three-stage protocol for implementing the Cubic transformation in public key cryptography systems.**

## 3. Asymmetric oblivious transfer arising from cubic transformation

Oblivious transfer was discussed by Rabin [5]. He used the square transformation which he had described in [2] and he discarded the cubic transformation as discussed in the introduction of this paper. We develop a condition for oblivious transfer for the cubic transformation and note that the use of higher power transformations will all provide asymmetric oblivious transfer. But in the next section, we also consider a method of asymmetric oblivious transfer using Diffie-Hellman key exchange protocol.

The oblivious transfer presented by Rabin uses the fact that if $x$ and $x_1$ are any two messages mapping to same cipher $c$, then gcd $(x - x_1, n) = p$ or $q$ with a probability of one-half.

We observe that we can increase the probability of oblivious transfer by using the special case of cubic transformation. The key to Kak's cubic transformation is that only one of the factors of $n$ is of the form $3k + 1$, thus only three messages mapping to one cipher. Here, if $x, x_1, x_2$ are the three messages mapping to the same cipher $c$, then both gcd $(x - x_1, n)$ and gcd$(x - x_2, n)$ yield the factors of $n$. Thus, the probability increases to two-thirds.

We can generalize this by considering transformations of the type $c = m^a \mod n$, where one of the factors of $n$ is of the form $ak + 1$ and $a$ is odd. Thus, only $a$ messages map to a single cipher and the probability of oblivious transfer becomes $\frac{a-1}{a}$. For example if $c = m^5 \mod n$ and only $p = 5k + 1$, then four out of five gcd's will yield the factors. Thus, the probability is $\frac{4}{5}$.

## 4. Diffie-Hellman based oblivious key exchange

The basic Diffie-Hellman key exchange protocol is useful in establishing a shared secret session key between the transmitter and the receiver. In its native form the protocol is as follows:

- Alice and Bob agree to use a prime number $p$ and a base $g$.
- Alice chooses a secret key $a$ and sends to Bob $g^a \mod p$.
- Bob chooses a secret key $b$ and sends to Alice $g^b \mod p$.
- Alice computes $(g^b \mod p)^a \mod p = S_{key}$.
- Bob computes $(g^a \mod p)^b \mod p = S_{key}$.

Since, all the calculations are performed modulo $p$, we can drop $\mod p$ from our notation. Thus, we can just write in general:

- Alice sends to Bob $g^a$ and Bob does $(g^a)^b = S_{key}$.

- Bob sends to Alice $g^b$ and Alice does $(g^b)^a = S_{key}$.

We now present a small modification to the above protocol which leads to an oblivious exchange of keys. The probability that both Alice and Bob have the same key is one-quarter. We present the use of square transformation here and hence the reader should keep in mind that two messages will map to same cipher.

The protocol we present is as follows:
- Alice and Bob agree to use a prime number $p$ and a base $g$.
- Alice and Bob agree upon one common cipher $c$ such that for any two numbers $x$ and $y$, we have $x^2 \mod p = y^2 \mod p = c$.
- Alice chooses a secret number $N_1$ and sends to Bob $g^{\sqrt{c}+N_1}$.
- Bob chooses a secret number $N_2$ and sends to Alice $g^{\sqrt{c}+N_2}$.
- Alice computes $\left(\dfrac{g^{\sqrt{c}+N_2}}{g^{\sqrt{c}}}\right)^{N_1} = S_{key1}$.
- Bob computes $\left(\dfrac{g^{\sqrt{c}+N_1}}{g^{\sqrt{c}}}\right)^{N_2} = S_{key2}$.

We notice that there are two possible values for $\sqrt{c}$, i.e. $\sqrt{c} = x$ and $\sqrt{c} = y$. Thus, $S_{key1} = S_{key2}$ with probability one-quarter.

**Example**: Let Alice and Bob decide upon $p = 19$ and base $g = 2$. They further choose to use $c = 9$.

We notice that $3^2 \mod 19 = 16^2 \mod 19 = 9$. Alice chooses a secret number $N_1 = 7$ and Bob chooses a secret number $N_2 = 11$. Thus, there are four possible cases that arise.

Suppose:
- Alice send to Bob: $2^{\sqrt{9}+7} = 2^{3+7} = 2^{10} \mod 19 = 17$.
- Bob sends to Alice: $2^{\sqrt{9}+11} = 2^{3+11} = 2^{14} \mod 19 = 6$.

*Case 1:*
- Alice does: $\left(\dfrac{2^{14}}{2^3}\right)^7 = (2^{11})^7 \mod 19 = 13$.
- Bob does: $\left(\dfrac{2^{10}}{2^3}\right)^{11} = (2^7)^{11} \mod 19 = 13$.

*Case 2:*
- Alice does: $\left(\dfrac{2^{14}}{2^{16}}\right)^7 = (2^{-2})^7 \mod 19 \neq 13$.
- Bob does: $\left(\dfrac{2^{10}}{2^3}\right)^{11} = (2^7)^{11} \mod 19 = 13$.

*Case 3:*
- Alice does: $\left(\dfrac{2^{14}}{2^{16}}\right)^7 = (2^{-2})^7 \mod 19 = x$.
- Bob does: $\left(\dfrac{2^{10}}{2^{16}}\right)^{11} = (2^{-6})^{11} \mod 19 \neq x$.

*Case 4:*
- Alice does: $\left(\dfrac{2^{14}}{2^3}\right)^7 = (2^{11})^7 \mod 19 = 13$.
- Bob does: $\left(\dfrac{2^{10}}{2^{16}}\right)^{11} = (2^{-6})^{11} \mod 19 \neq 13$.

We see that each party gets the same secret key as the other with probability one-quarter and neither party knows if the other has the same session key.

The Diffie-Hellman-based oblivious key exchange can also be used for higher exponents that would lead to other probability values. For example, if we use the cubic transformation the probability would decrease to one-ninth.

## 5. Conclusion

Rank information can also help in the case of Rabin's square transformation [2]. But since every prime number is odd, each cipher will have four possible messages modulo a composite number $n$. Therefore, we need two extra bits to send the rank information, which is slightly inefficient compared to cubic transformation as there is only three-messages-to-one-cipher mapping and hence one bit less for transmission.

Also, we have introduced a new method of oblivious transfer using Diffie-Hellman key exchange. As seen this new type of oblivious transfer is asymmetric in nature and gives rise to probabilities other than one-half. The asymmetric nature of oblivious transfer may be useful in e-commerce applications where a party wishes to covey less than half the information before a payment is made.